\documentclass{PoS}
\usepackage{subfigure}
\title{VLBA imaging of radio-loud BAL QSOs}

\ShortTitle{VLBA imaging of radio-loud BAL QSOs}

\author{\speaker{F.M.~Montenegro Montes}$^{abc}$, K.-H.~Mack,$^a$ 
        C.R.~Benn,$^d$ R.~Carballo,$^e$ D.~Dallacasa,$^{fa}$ 
	J.I.~Gonz\'alez-Serrano,$^g$ J.~Holt$^h$ and 
	F.~Jim\'enez-Luj\'an$^{gi}$\\
        \llap{$^a$} INAF - Istituto di Radioastronomia, Via P. Gobetti 101, I-40129 Bologna, Italy\\
        \llap{$^b$} Dpto. de Astrof\'isica, Universidad de La Laguna, Avda. Astrof\'isico Fco. S\'anchez s/n, E-38209 La Laguna, Spain\\
        \llap{$^c$} Instituto de Astrof\'isica de Canarias, C/ Via L\'actea s/n, E-38200 Tenerife, Spain\\
	\llap{$^d$} Isaac Newton Group, Apartado 321, E-38700 Santa Cruz de La Palma, Spain\\
        \llap{$^e$} Dpto. de Matem\'atica Aplicada y Ciencias de la Computaci\'on, Univ, de Cantabria, ETS Ingenieros de Caminos
                    Canales y Puertos, \\ 
                    Avda. de los Castros s/n, E-39005 Santander, Spain\\
        \llap{$^f$} Dipartimento di Astronomia, Universit\`a di Bologna, Via Ranzani 1, I-40127 Bologna. Italy\\
        \llap{$^g$} Instituto de F\'isica de Cantabria (CSIC-Universidad de Cantabria), Avda. de los Castros s/n, E-39005 Santander, Spain\\
        \llap{$^h$} Leiden Observatory, Leiden University, P O Box 9513, NL-2300 RA Leiden, The Netherlands\\
        \llap{$^i$} Dpto. de F\'isica Moderna,  Univ. de  Cantabria, Avda de los Castros s/n, E-39005 Santander, Spain

        E-mail: \email{fmm@ira.inaf.it}}

\abstract{Broad Absorption Line Quasars (BAL QSOs) have been found to be associated with extremely compact radio 
sources. These reduced dimensions can be either due to projection effects or these objects might actually be 
intrinsically small. Exploring these two hypotheses is important to understand the nature and origin of the
BAL phenomenon because orientation effects are an important discriminant between the different models proposed 
to explain this phenomenon. In this work we present VLBA observations of 5 BAL QSOs and discuss their 
pc-scale morphology.}

\FullConference{The 9th European VLBI Network Symposium on The role of VLBI in the Golden Age for Radio Astronomy and EVN Users Meeting\\
		 September 23-26, 2008\\
		 Bologna, Italy}

\begin{document}

\section{Introduction}

The presence of Broad Absorption Lines (BALs) in the optical spectra of some quasars is interpreted as a 
footprint of powerful quasar winds accelerated by the central engine up to velocities of $\sim$0.2c.
Broad Absorption Line Quasars (BAL QSOs) have been revealed to be associated with extremely compact radio 
sources with projected linear sizes smaller than 1 kpc. To explain this, one possibility is that these 
could be intrinsically small objects but they might also be compact due to projection effects. The first 
hypothesis would find a natural explanation within the so-called \emph{evolutionary scenario} proposed to 
interpret the nature of BAL QSOs. A second option would suggest that BAL QSOs could be oriented in a 
particular way, the basis of some \emph{orientation scenarios} that explain in this way the detection of 
the BAL phenomenon in a subset of $~$15-20\% of all QSOs.\\

In [1] it was shown that radio BAL QSOs share several properties with the class of
Gigahertz-Peaked Spectrum (GPS) sources, which are radio sources going through the first stages of their 
evolution. VLBI observations are crucial to reveal the pc-scale structure of these objects and help 
understand their geometrical disposition. For this purpose we present here two-frequency VLBA observations 
of 5 radio BAL QSOs.

\section{Sample and observations}

The list of observed objects can be found in Table \ref{tableobsVLBA}, with columns being the observation date, 
radio coordinates in J2000, redshift and flux densities at 4.8 and 8.3 GHz from previous continuum Effelsberg 
observations. Four of the BAL QSOs belong to the sample presented in [1], and we have also observed the radio 
BAL QSO 1537$+$58. Details on the radio spectra and polarisation properties of these objects can be found 
in [1].

\begin{table}[h]
 \centering
 \small
\caption{List of 5 BAL QSOs observed with VLBA}\label{tableobsVLBA}
\begin{tabular}{ccrrccc}
 \hline
   ID	 & Obs. Date &	  RA	  &	DEC          & z  & S$_{4.8\, \rm{GHz}}$& S$_{8.3\, \rm{GHz}}$ \\
         &           &	(J2000)   &   (J2000)        &	  & (mJy)               & (mJy)                \\
\hline
0135$-$02& 21-10-2007  & 01 35 15.243 & $-$02 13 49.37 & 1.820 &  27.2 $\pm$ 0.6	&  31.9 $\pm$ 0.8     \\
0837$+$36& 06-11-2007  & 08 37 45.589 & $+$36 41 45.60 & 3.416 &  30.2 $\pm$ 2.3	&  12.1 $\pm$ 0.8     \\
1159$+$01& 23-12-2007  & 11 59 44.827 & $+$01 12 06.98 & 1.989 & 137.8 $\pm$ 1.7	& 158.0 $\pm$ 2.0     \\
1537$+$58& 25-11-2007  & 15 37 29.549 & $+$58 32 24.80 & 3.059 &  26.5 $\pm$ 0.4	&  40.6 $\pm$ 0.8     \\
%1603$+$30&           & 16 03 54.159 & $+$30 02 08.88 & 2.028 &  26.1 $\pm$ 0.7	&  19.1 $\pm$ 0.5     \\
1624$+$37& 23-01-2005  & 16 24 53.481 & $+$37 58 06.66 & 3.377 &  23.3 $\pm$ 1.1	&  15.0 $\pm$ 0.1     \\
%14 13 18.865 & $+$45 05 22.98& 141.8  & 1.5     & 173.3  & 1.5     \\
\hline
\end{tabular}
\end{table}

Two receivers were used, centred on the C and X bands (at about 5.0 and 8.4 GHz respectively). A 
minimum of two frequencies allows us to have different resolutions to look for individual components and 
to study the spectral indices and the polarisation properties of sub-components. Details on the observational 
setup, calibration, imaging and further analysis of the VLBA data will be presented in a forthcoming paper 
currently in preparation.

\section{VLBA maps and discussion}

\begin{figure}[h!!!!]
\begin{center}{
\mbox{
   \subfigure{
     \hspace{-0.5cm}
      \includegraphics[width=6.5cm]{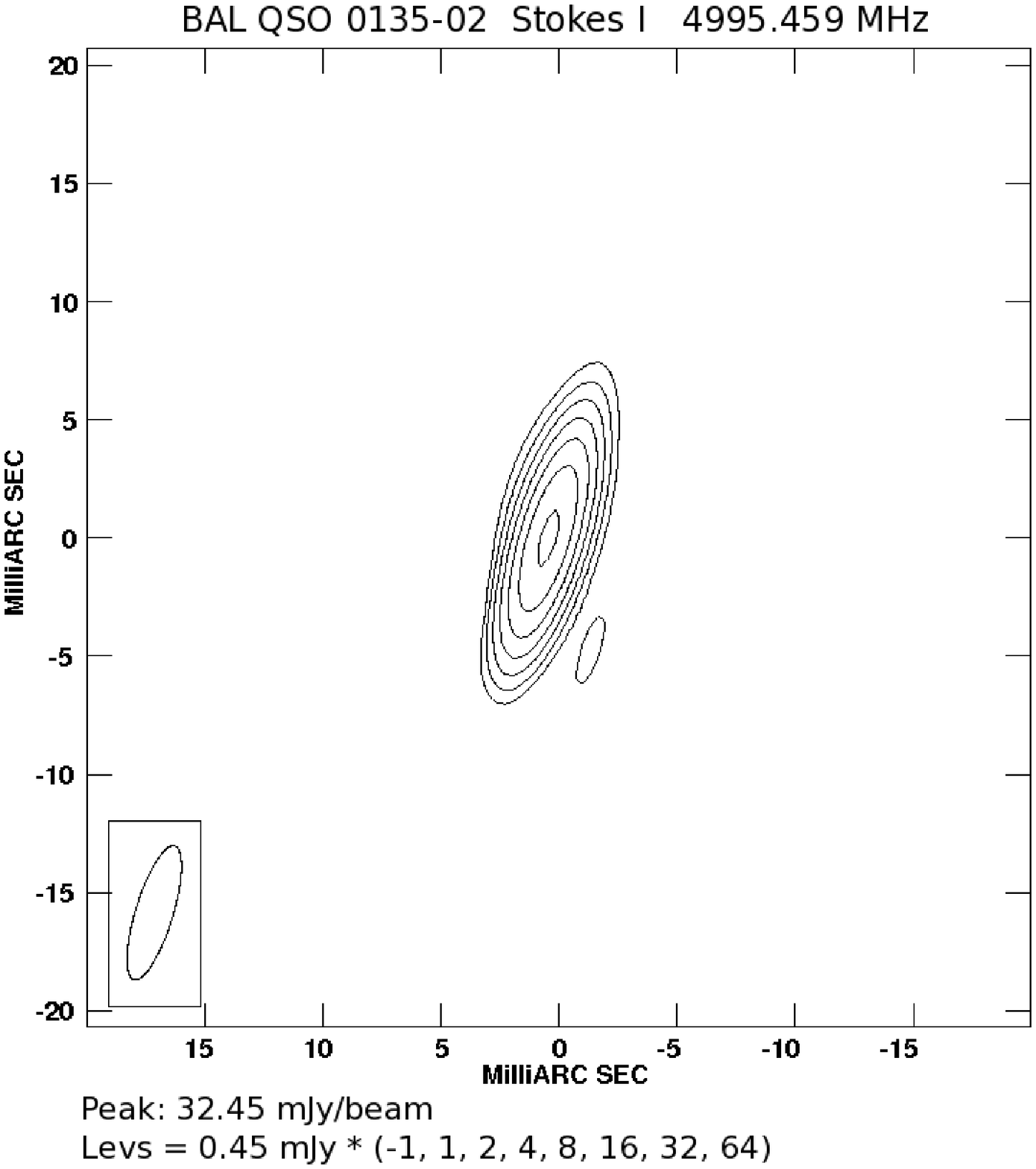}}
  \hspace{0.2cm}
   \subfigure{
     \hspace{-0.5cm}
      \includegraphics[width=6.2cm]{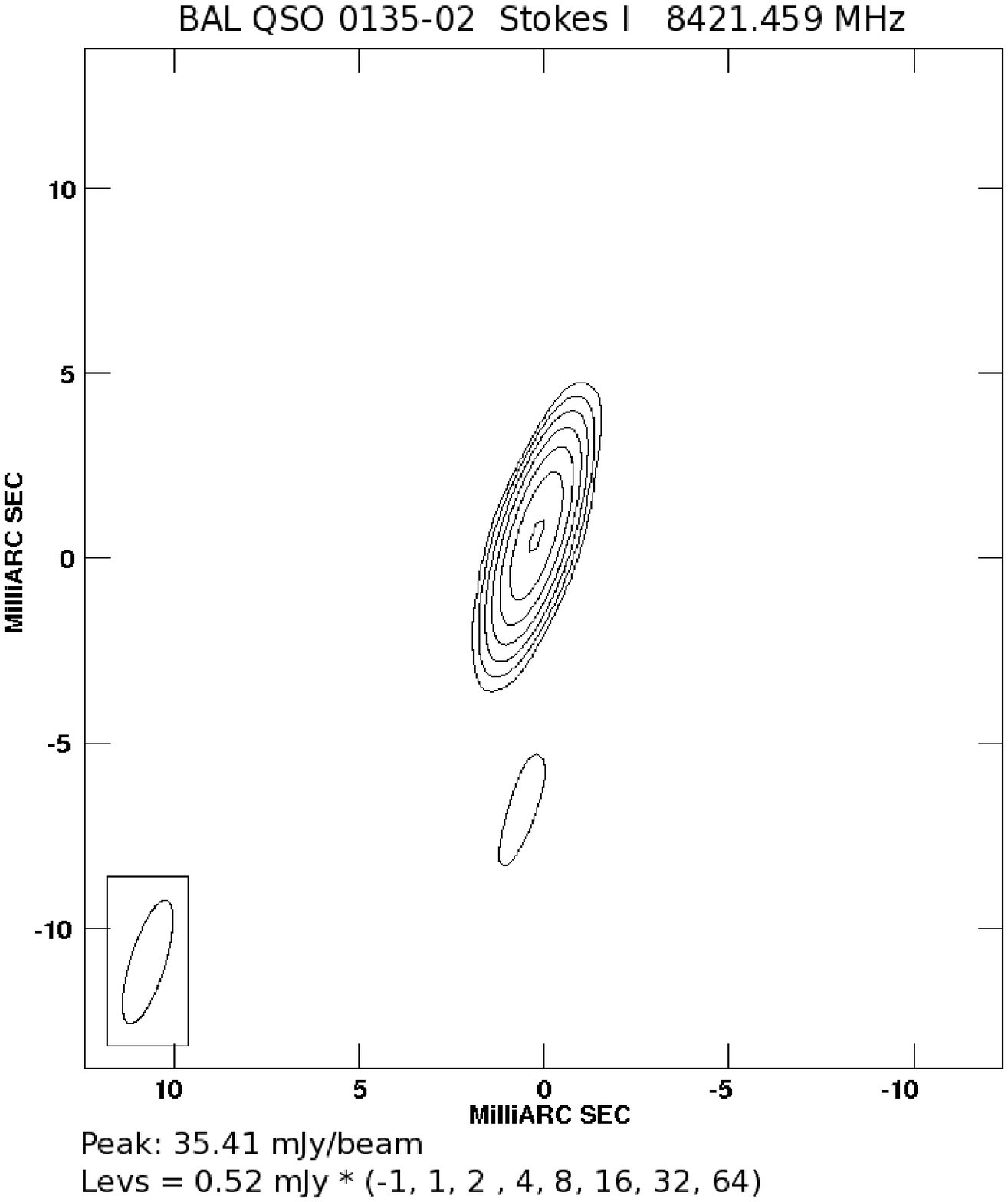}}
 }
\mbox{
   \subfigure{
     \hspace{-0.5cm}
      \includegraphics[width=6.5cm]{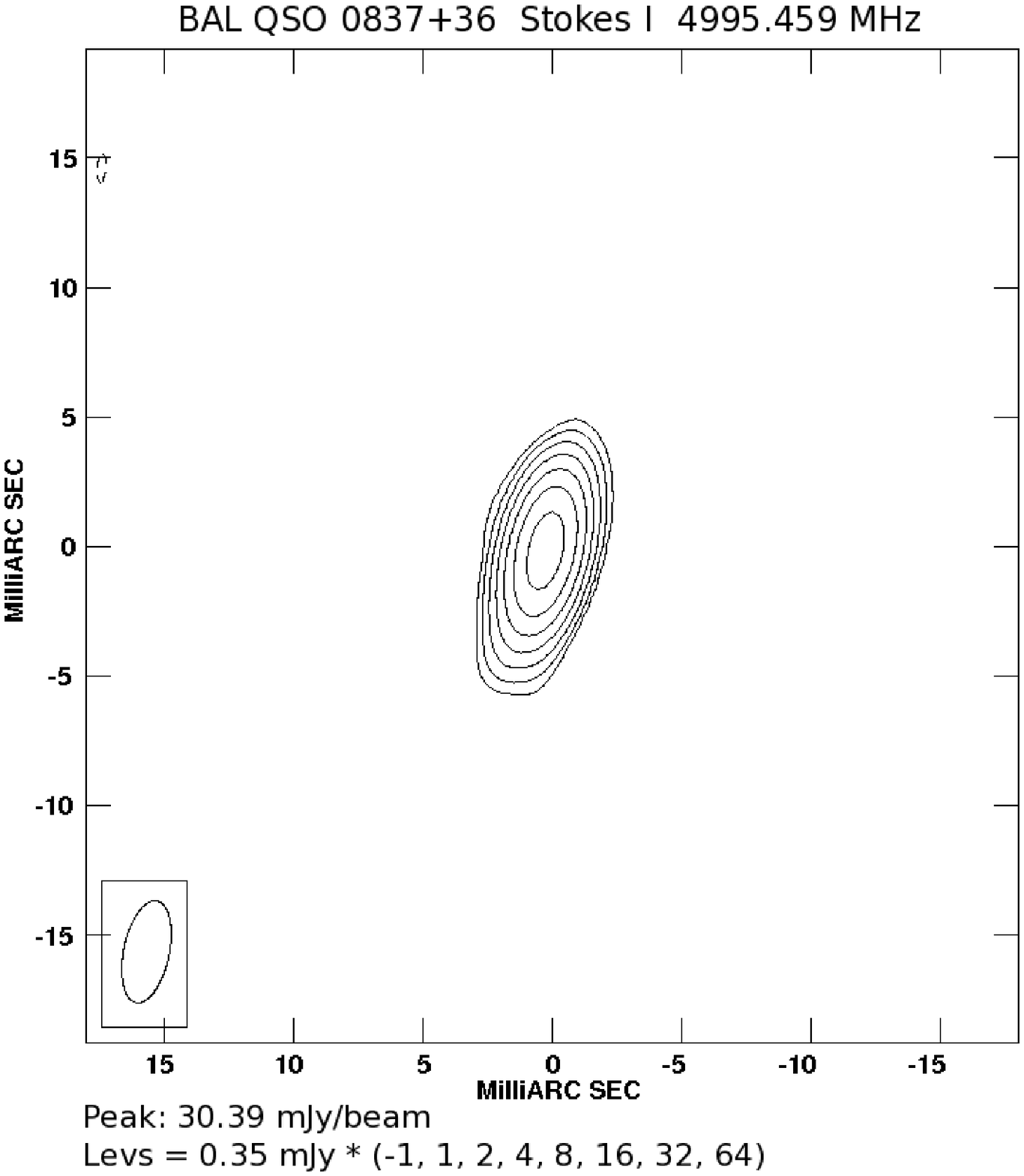}}
  \hspace{0.2cm}
   \subfigure{
     \hspace{-0.5cm}
      \includegraphics[width=6.5cm]{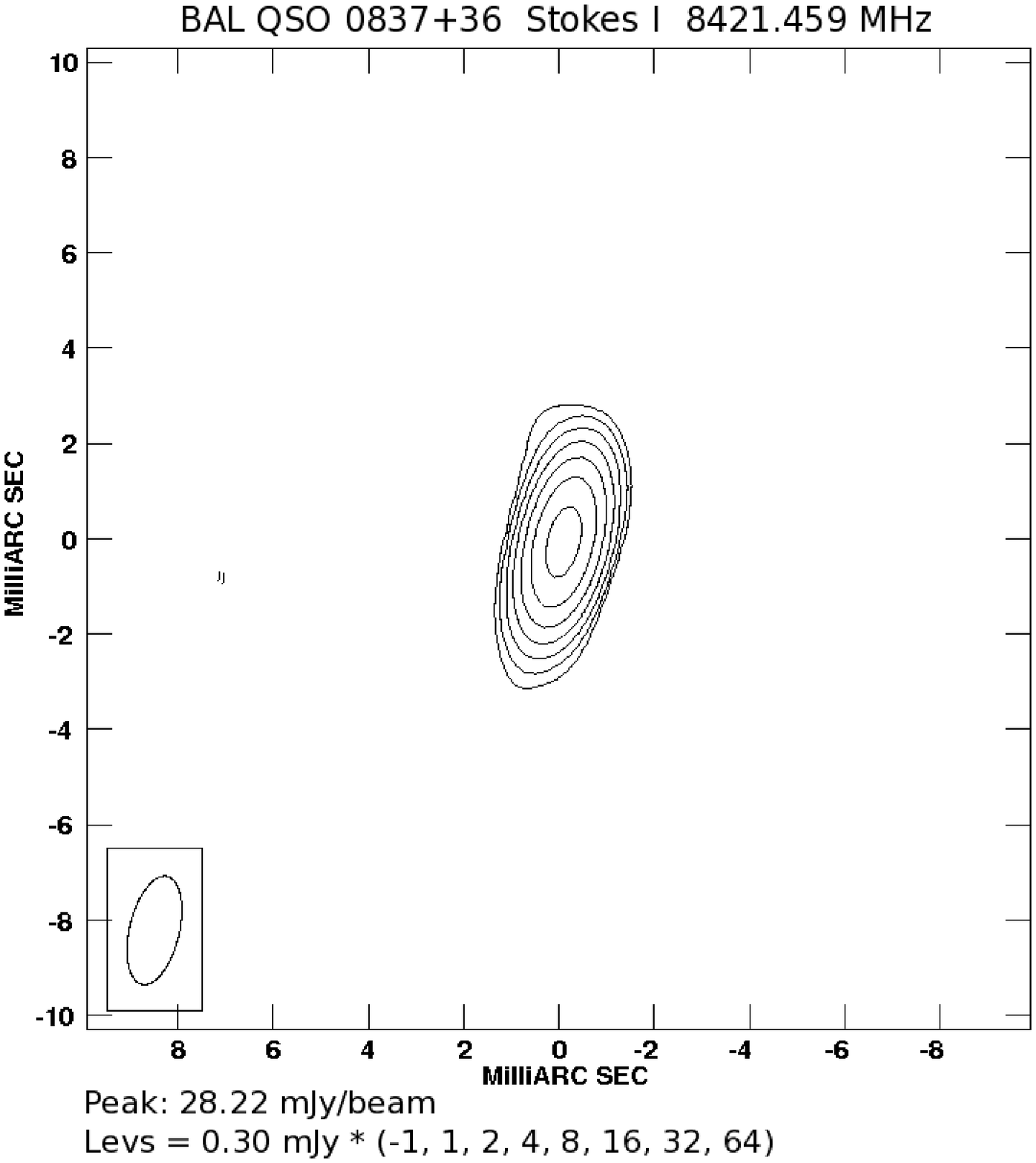}}
 }
 
{\footnotesize \textit{\caption{VLBA maps at 5 GHz (left) and 8 GHz (right) showing the pc-scale structure 
of BAL QSOs 0135$-$02 (top) and 0837+36 (bottom). The synthesised beam size is shown in the lower left corner 
of the map. Contours are in logarithmic scale according to the legend.}\label{VLBA_1}}}
}
\end{center}
\end{figure}

The VLBA maps of BAL QSOs 0135$-$02 and 0837+37 are shown in Figure \ref{VLBA_1}, where the peak flux 
densities are also noted. These two sources are unresolved with the VLA at 8.4 GHz, and they still appear 
point-like at 
the milliarc\-second-scale. Gaussian fits at 8.4 GHz give angular extensions of 3.5 $\times$ 0.9 mas and 2.3 
$\times$ 1.0 mas for 0135$-$02 and 0837+37, respectively. These translate into linear dimensions of about 
30 $\times$ 8 pc and 19 $\times$ 8 pc for these two sources, respectively.

When compared to our previous Effelsberg and VLA measurements, there are slight increases of about 20\% and 
10\% in the flux densities of 0135$-$02, at 5.0 and 8.4 GHz, respectively. This might suggest moderate 
variability, and probably excludes the possibility of a fainter extended component not recovered in the VLBA 
maps. However, for 0837+36, the 8.4-GHz flux density would have increased about 130\% in 2.3 years (observer's 
time), with a variability significance $\sigma_{Var}$ = $\left( S_{2}-S_{1} \right)$/$\sqrt{\sigma_{1}^2 + 
\sigma_{2}^2}$ = 5.2, while at 5.0 GHz both the Effelsberg and the VLBA measurements give similar values.
If the variability at 8.4 GHz is real and not due to a calibration error, the radio spectrum of this source 
has considerably flattened between these two frequencies.

\begin{figure}[h!!!!]
\begin{center}{
\mbox{
   \subfigure{
     \hspace{-0.5cm}
      \includegraphics[width=6.5cm]{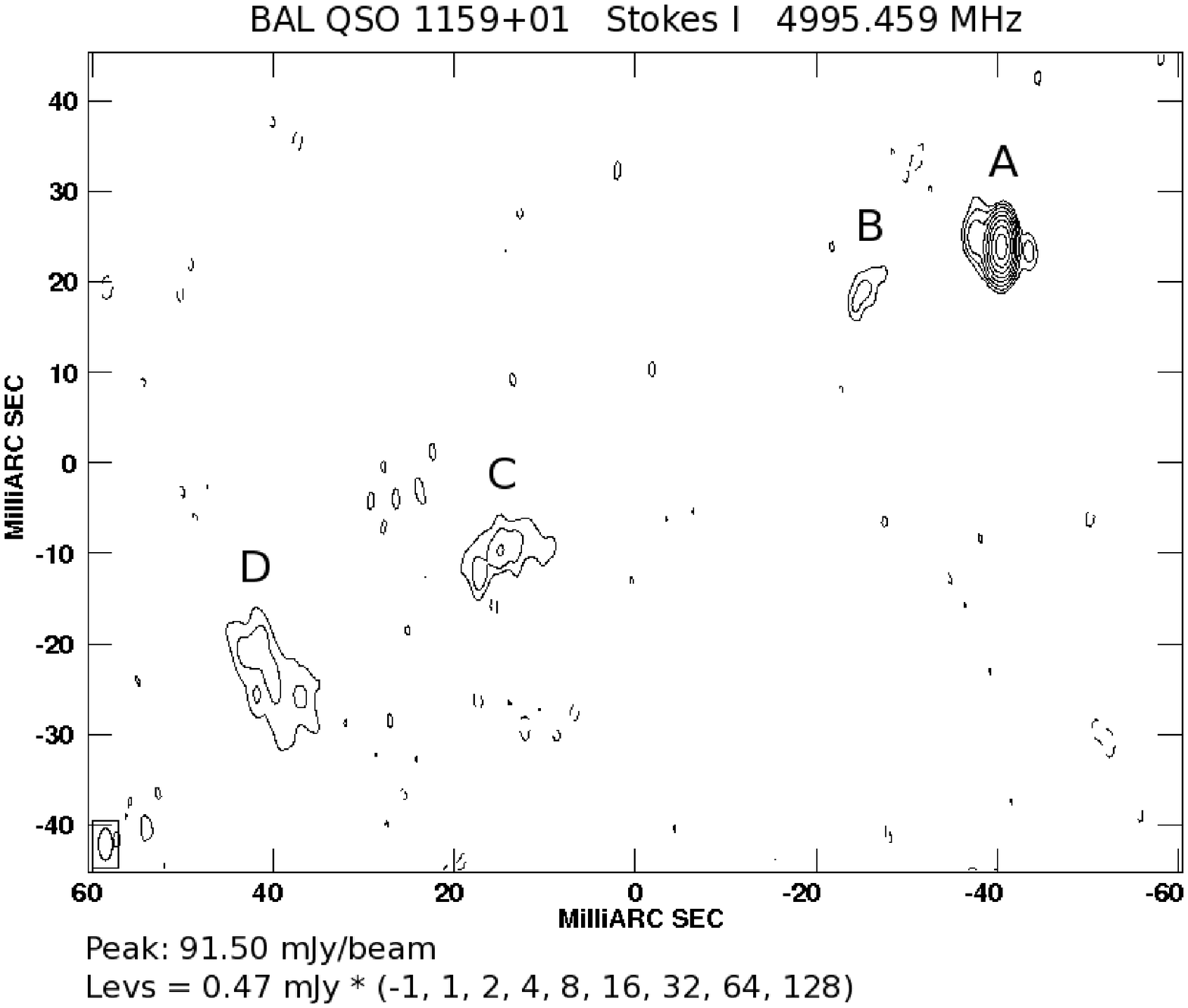}}
  \hspace{0.2cm}
   \subfigure{
     \hspace{-0.5cm}
      \includegraphics[width=7.2cm]{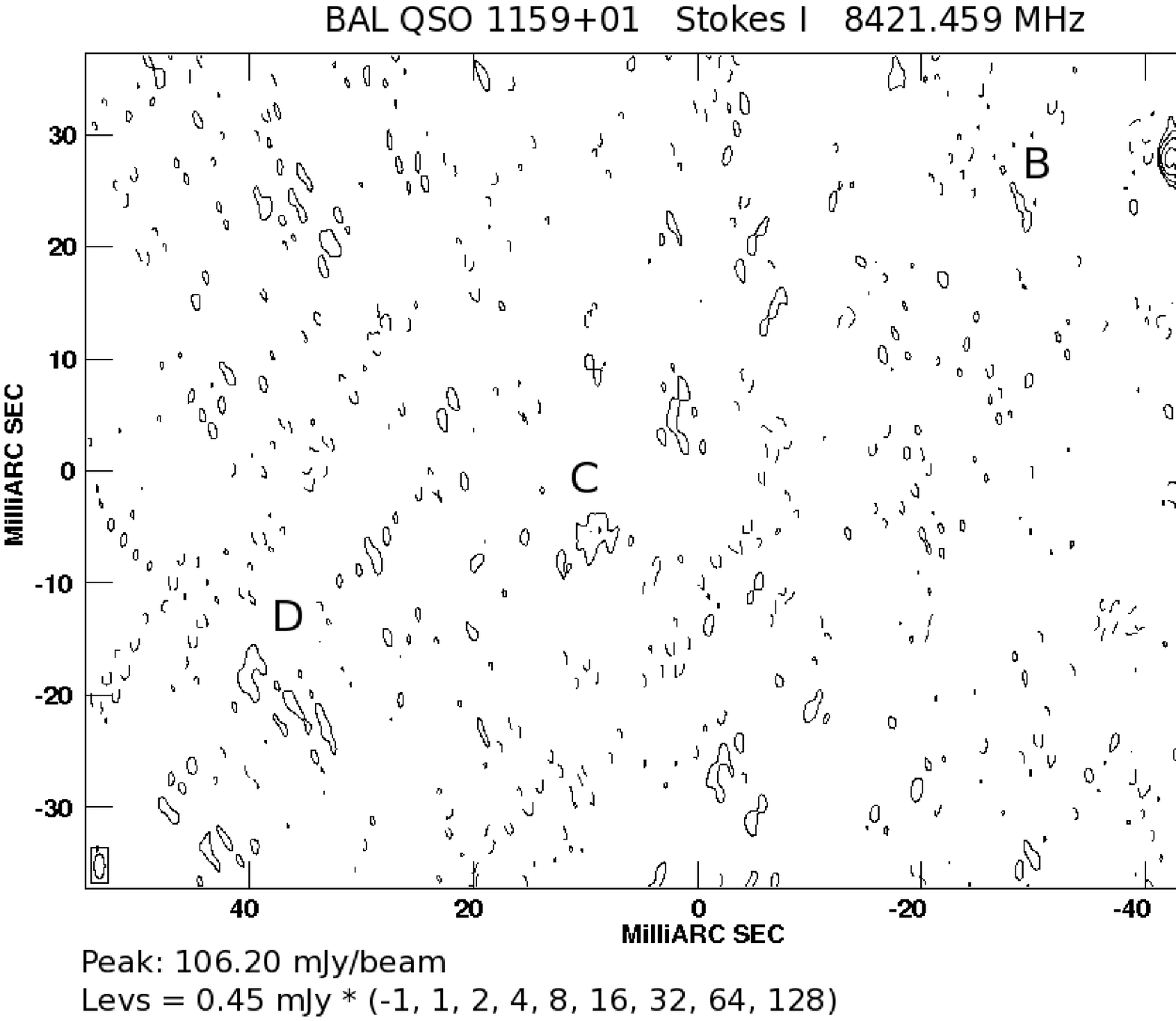}}
 }
\mbox{
   \subfigure{
     \hspace{-0.5cm}
      \includegraphics[width=6.5cm]{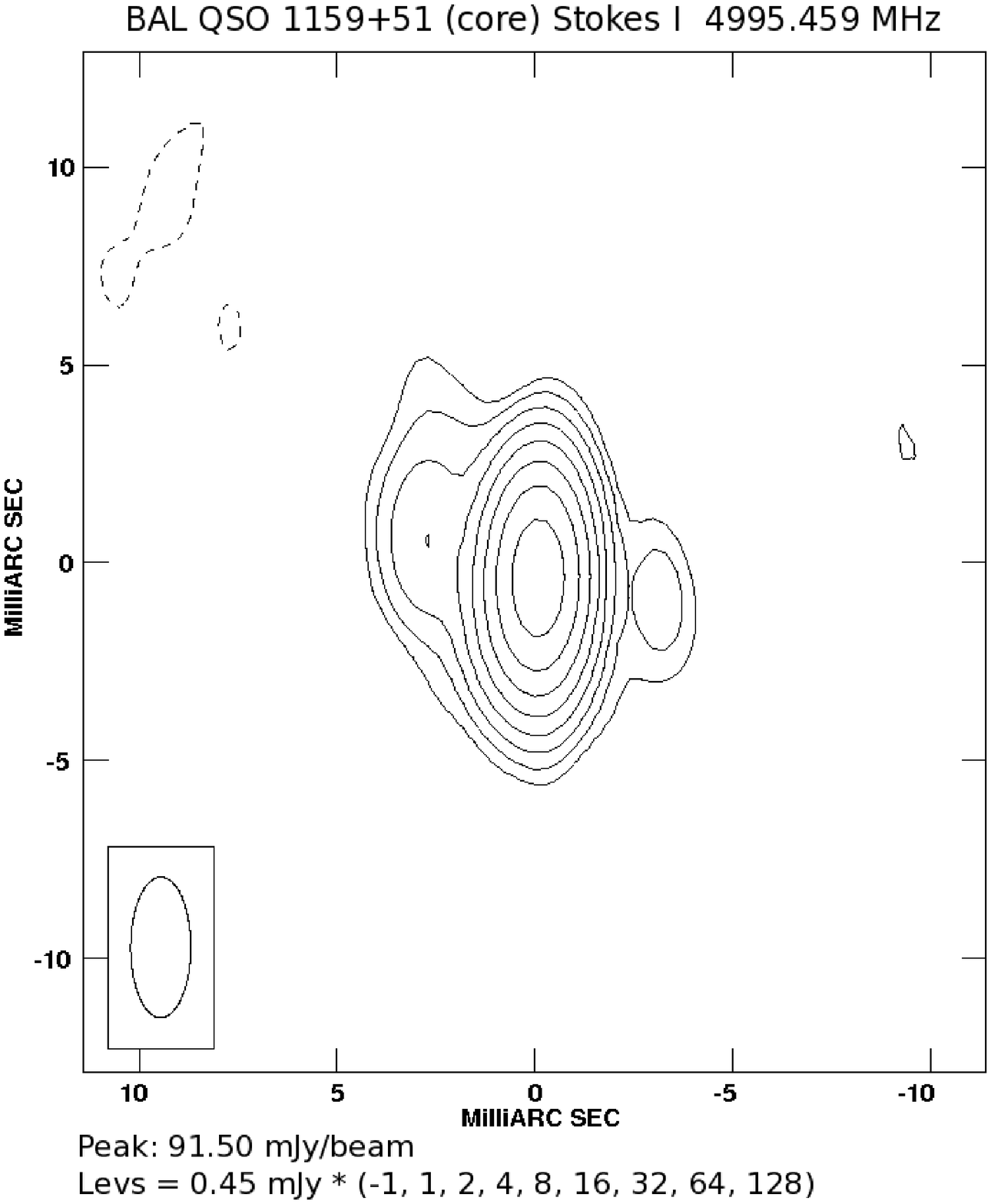}}
  \hspace{0.2cm}
   \subfigure{
     \hspace{-0.5cm}
      \includegraphics[width=7.0cm]{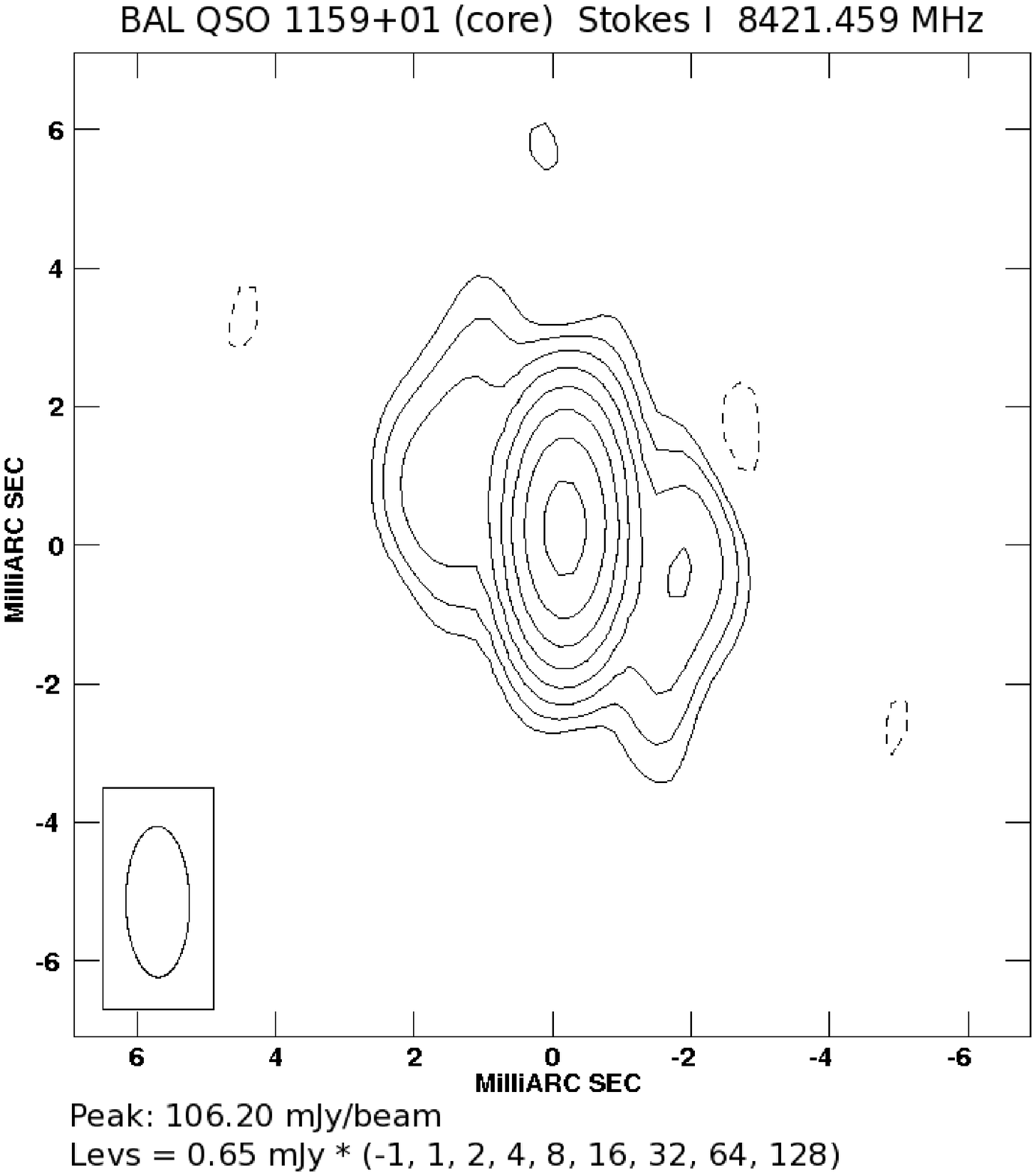}}
 }
 
{\footnotesize \textit{\caption{VLBA maps at 5 GHz (left) and 8 GHz (right) of BAL QSO 1159+01. The upper
plots show the largest scale structure, while the lower plots detail the nuclear region. The synthesised 
beam size is shown in the lower left corner of the map. Contours are in logarithmic scale according to the 
legend.}\label{VLBA_2}}}
}
\end{center}
\end{figure}

In Figure \ref{VLBA_2} we see the extended structure of the interesting radio-loud BAL QSO 1159+01.
A compact and bright component (A) can be seen associated with the radio core and also some faint extended 
emission which extends about 100 mas towards the SE from the central core. The faintness of this emission and 
the absence of an active hot-spot suggest this to be the remainder of a jet that has ceased its activity. 
An estimate of the extension of the diffuse emission has been determined measuring the distance between the A 
and D components, yielding 100 mas, which corresponds to a projected linear size of 0.85 kpc. If there is no 
significant variability, no additional flux density is expected at longer distances from the core because the 
previously measured Effelsberg flux density is completely recovered when adding up the flux densities of the 4 
components (A to D). The faintness of components B, C and D in the 8.4-GHz map is consistent with a steep 
spectral index, which is expected if the emission comes from relative relaxed particles in a non-active region.
Also at this frequency the summed flux density from all 4 components approximately equals the flux density from 
the VLA maps.\\

\begin{figure}[t]
\begin{center}{
\mbox{
   \subfigure{
     \hspace{-0.5cm}
      \includegraphics[width=6.5cm]{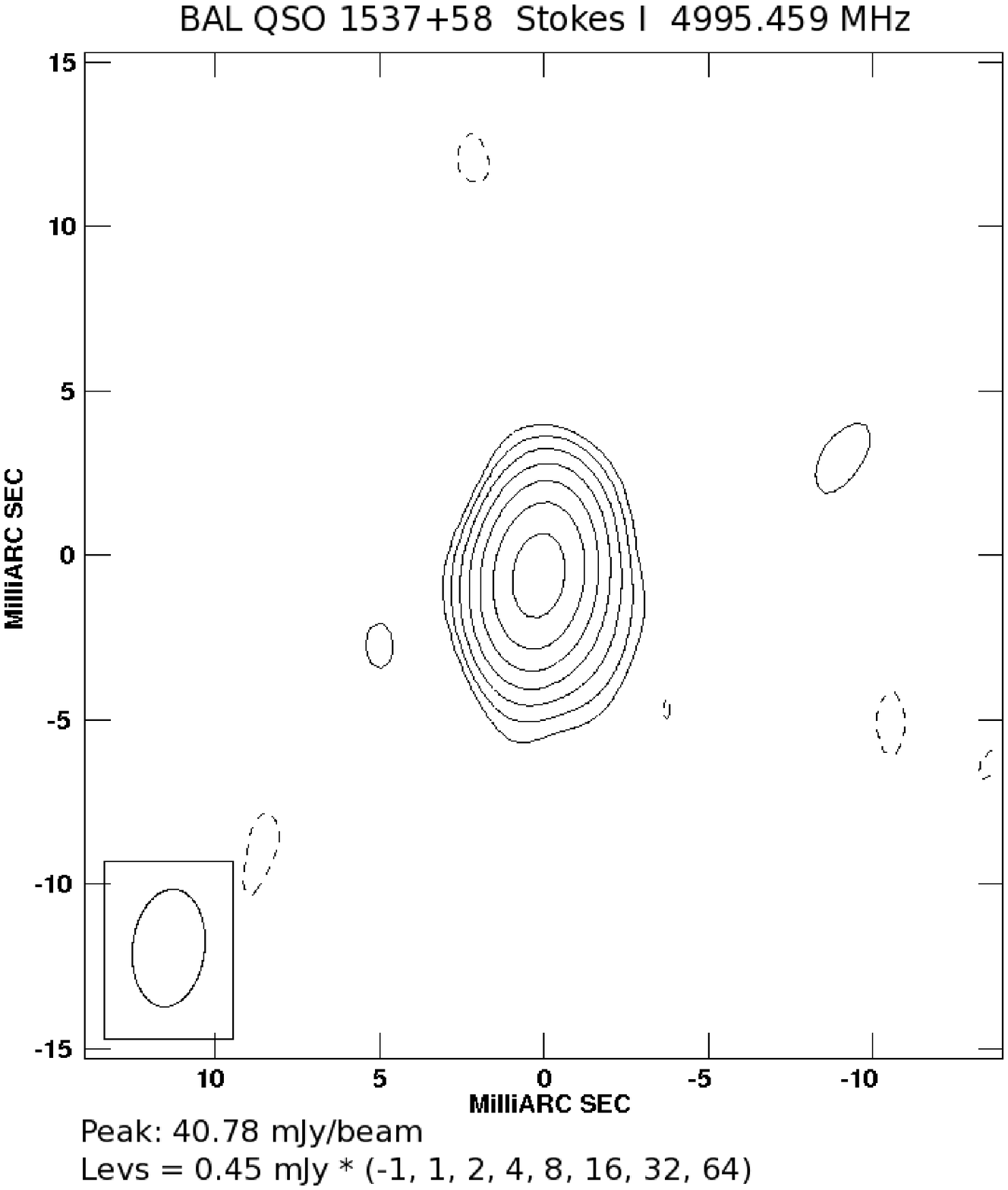}}
  \hspace{0.2cm}
   \subfigure{
     \hspace{-0.5cm}
      \includegraphics[width=6.5cm]{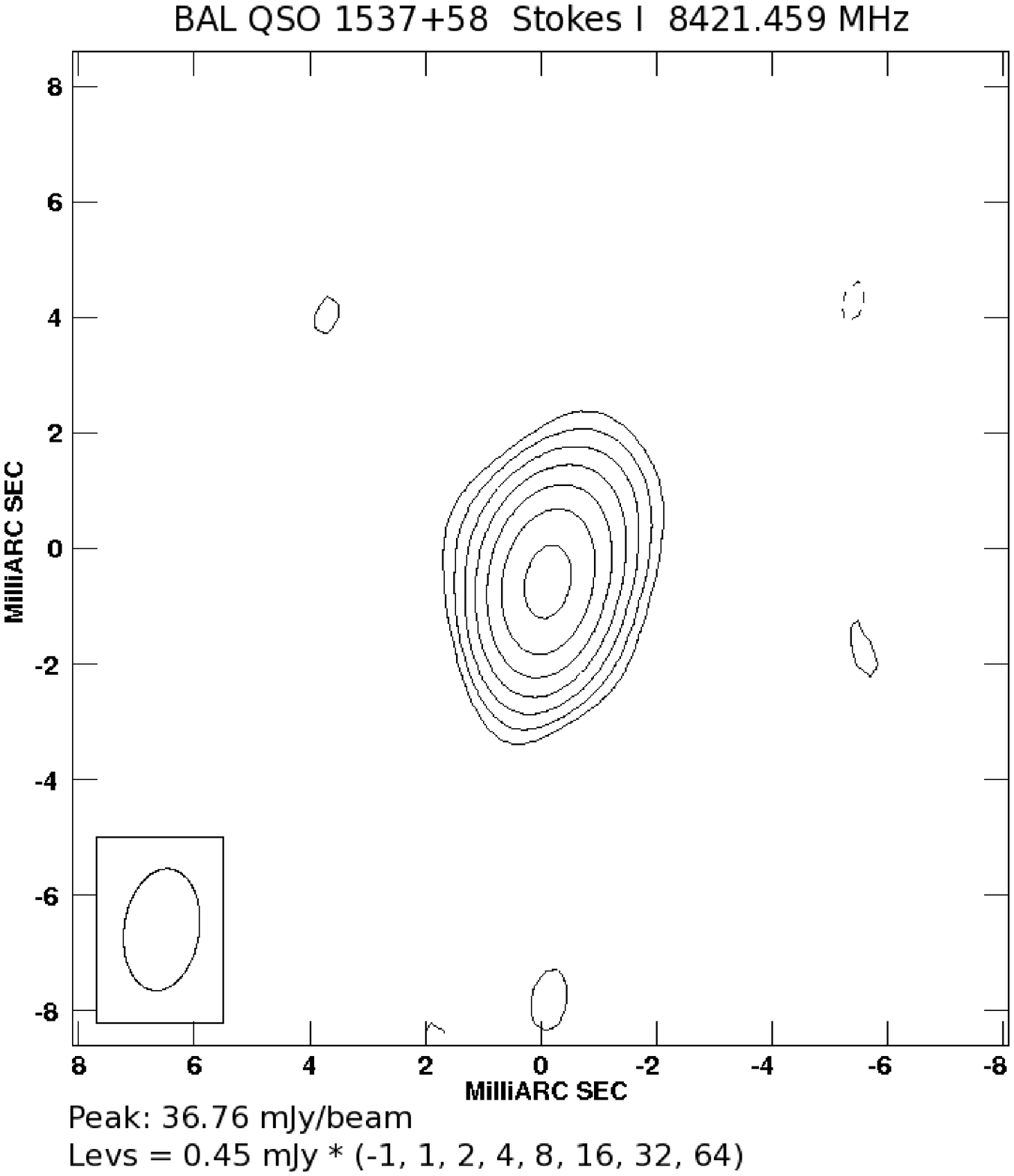}}
 }
\mbox{
   \subfigure{
     \hspace{-0.5cm}
      \includegraphics[width=6.5cm]{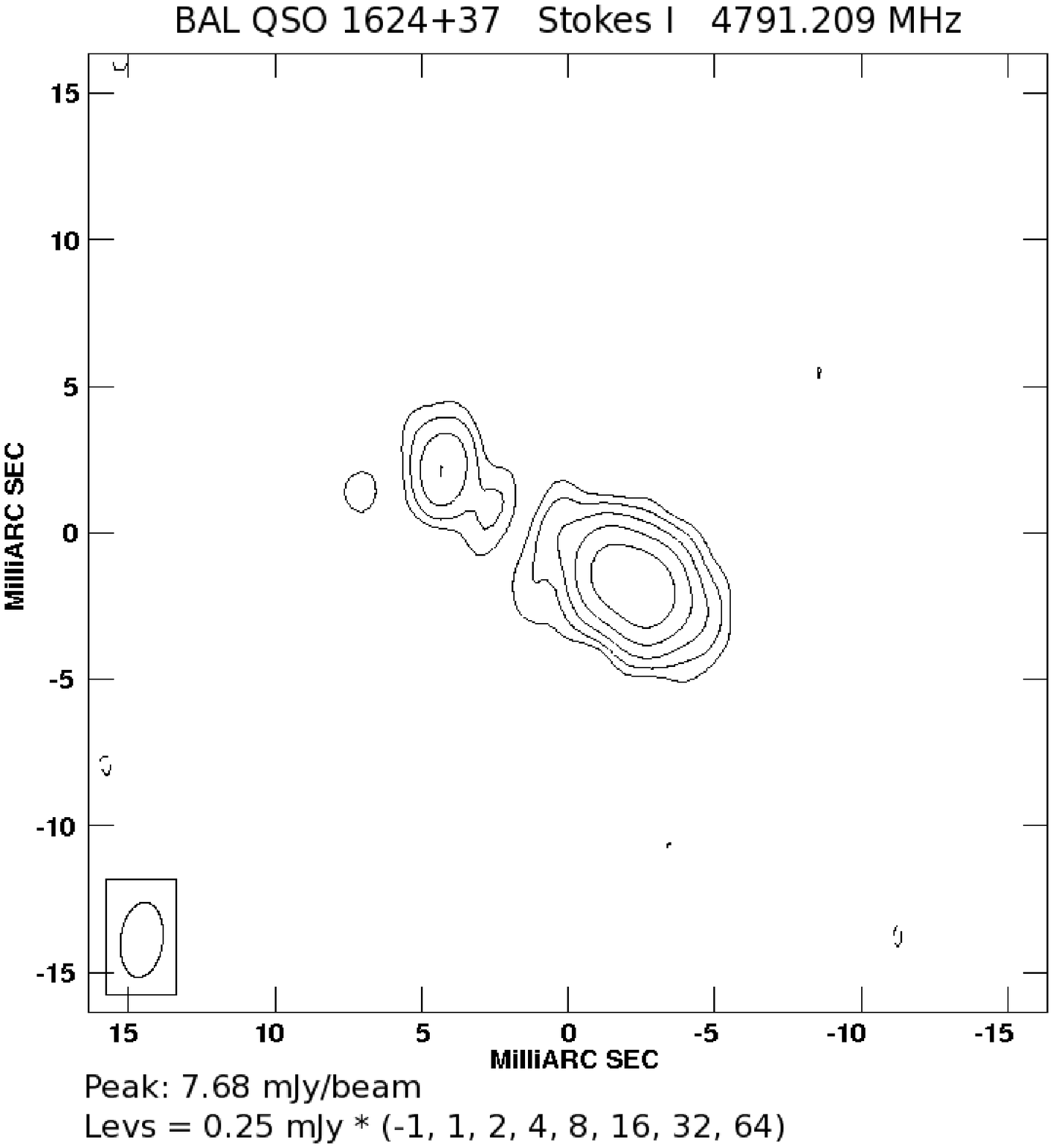}}
  \hspace{0.2cm}
   \subfigure{
     \hspace{-0.5cm}
      \includegraphics[width=6.7cm]{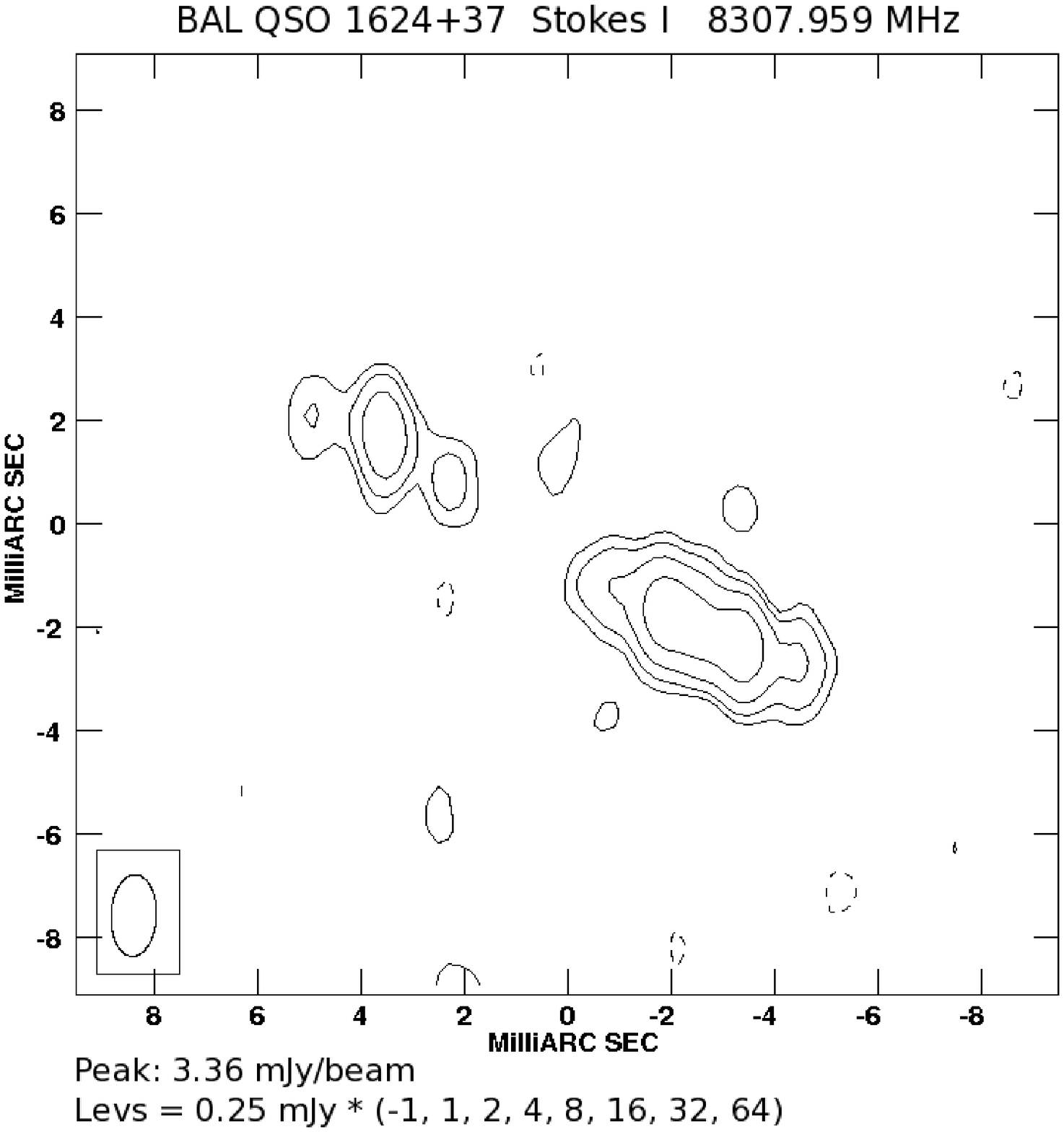}}
 }
 
{\footnotesize \textit{\caption{VLBA maps at 5 GHz (top) and 8 GHz (bottom) showing the pc-scale structure 
of BAL QSO 0135-02. The synthesised beam size is shown in the lower left corner of the map. Contours are 
in logarithmic scale according to the legend.}\label{VLBA_3}}}
}
\end{center}
\end{figure}

Another remarkable characteristic of 1159+51 is the presence of two symmetric extensions very close to
the central core (see the bottom of Figure \ref{VLBA_2}). At both 5.0 and 8.4 GHz a central brighter core and 
two symmetric components at about 2 arcsec from the centre can be seen. One of them is located towards the NWW 
and the other in the opposite direction. The first one is slightly brighter than the other at 5 GHz, but at 
8 GHz they have similar flux densities as can be seen in the contour map. This might be the origin of a new jet 
coming from the central nucleus. These two jets would not be aligned but both directions form an angle of about 
50$^{\circ}$.\\

Finally, Figure \ref{VLBA_3} shows the VLBA maps of 1537+58 and 1624+37. The first one looks quite compact
and is just barely resolved at both frequencies. As in the cases of 0135$-$02 and 0837+36 some variability 
can reveal signs of beaming, with a flux density increase of $\sim$50\% at 5 GHz and $\sigma_{Var}$ = 3.5.
The unusual BAL QSO 1624+37 shows a core-jet structure clearly visible at both frequencies. There is a more 
compact component (core) with flat spectral index and a more extended component towards the SE direction (jet)
with steeper spectral index. The separation between the emission peaks of the two components is 8 mas. The core 
seems to have a very faint component at 1-2 mas towards the jet direction that could be the emission from the 
base of the jet.

\section{Conclusions}

Our contribution increases the number of VLBI observations of BAL QSOs published in the literature. 
A total of 17 of these sources can be found combining this work with [2], [3] and [4]. A variety of
geometries are found among these sources: 7/17 are unresolved with VLBI baselines, 7/17 show a core-jet 
structure and 3/17 show a central bright core and two fainter extensions in approximately opposite directions,
similar to Compact Symmetric Objects (CSOs). This variety of geometry is difficult to reconcile with 
simple orientation scenarios trying to explain the nature of BAL QSOs.

\end{document}